\documentclass{PoS}

\title{$B$ spectroscopy  with NRQCD and HISQ}

\ShortTitle{$B$ spectroscopy}

\author{\speaker{Eric B. Gregory}\thanks{Current address: Department of Physics,
University of Cyprus, P.O. Box 20357, 1678 Nicosia, CYPRUS}\\
        University of Glasgow \\
        E-mail: \email{egregory@ucy.ac.cy}}

\author{Christine.T.H. Davies\\
        University of Glasgow}
\author{Eduardo Follana\\
         Universidad de Zaragoza}
\author{Elvira G\'{a}miz\\
        University of Illinois}
\author{Iain Kendall\\
        University of Glasgow}
\author{G. Peter Lepage\\
        Cornell University}
\author{Heechang Na\\
        The Ohio State University}
\author{Junko Shigemitsu\\
        The Ohio State University}
\author{Kit Y. Wong \\
        University of Glasgow}

\abstract{Using NRQCD $b$ quarks and HISQ light, strange and charm quarks 
we have calculated $B$ meson masses and $B^*$-$B$ splittings. We quote 
results for a range of lattice spacings and sea quark masses, enabling 
controlled extrapolation to the physical point. Since the $b$ quark masses 
and lattice spacing are fixed from the $\Upsilon$ and other meson masses, 
this allows accurate, 
parameter-free tests of $B$, $B_s$, and $B_c$ masses against experiment. 
We can also predict the mass of the $B^*_c$ meson.}

\FullConference{The XXVII International Symposium on Lattice Field Theory 
- LAT2009\\
		 July 26-31 2009\\
		 Peking University, Beijing, China}

\newcounter{saveeqn}
\newcommand{\romeqn}{\setcounter{saveeqn}{\value{equation}}%
  \renewcommand{\theequation}{\Roman{equation}}
}
\newcommand{\reseteqn}{\setcounter{equation}{\value{saveeqn}}%
  \renewcommand{\theequation}{\arabic{equation}}}

\begin{document}

\section{Introduction}
The $B$ meson sector is a compelling target for lattice calculations for a 
variety of reasons. There are a variety of so-called ``gold-plated'' states --- 
those states which are narrow and hadronically stable, as well as being 
experimentally accessible. In calculating these properties of states on the 
lattice there are no free parameters; $m_\pi$, $m_K$, $m_{\eta_c}$ and 
$m_ {\Upsilon}$ calibrate the masses of the light, strange, charm and bottom 
quarks respectively, and $\Upsilon$ splittings and other meson masses 
calibrate the lattice 
spacing\cite{Davies:2009ts,Follana:2006rc,Follana:2007uv}. 

Precision $B$ meson spectroscopy is a key ingredient in precision
calculation of decay constants and form factors, ingredients in 
CKM matrix element determination and testing of the standard model.

\section{Simulation Methods}
We use five different ensembles of gauge configurations with $2+1$ flavors of 
dynamical ASQTAD sea quarks, generated by the MILC collaboration. The 
ensembles, listed in Table \ref{tab:params}, represent three lattice spacings,
labeled very-coarse, coarse, and fine.

\begin{table}
\begin{center}
\begin{tabular}{lllllllll}
\hline
\hline
Set & $\beta$ & $r_1/a$ & $au_0m_{0l}^{asq}$ & $au_0m_{0s}^{asq}$ & $L/a$ & $T/a$ & $N_{conf}\times N_{t}$ \\
\hline
1 & 6.572 & 2.152(5) & 0.0097 & 0.0484 & 16 & 48 & $624 \times 2$\\
2 & 6.586 & 2.138(4) & 0.0194 & 0.0484 & 16 & 48 & $628 \times 2$\\
\hline
3 & 6.760 & 2.647(3) & 0.005 & 0.05 & 24 & 64 & $507 \times 2$ \\
4 & 6.760 & 2.618(3) & 0.01 & 0.05 & 20 & 64 & $589 \times 2 $ \\
\hline
5 & 7.090 & 3.699(3) & 0.0062 & 0.031 & 28 & 96 & $530 \times 4$ \\
\hline
\hline
\end{tabular}
\end{center}
\caption{\label{tab:params}Ensembles (sets) of MILC configurations used with gauge coupling $\beta$, 
size $L^3 \times T$ and sea 
masses ($\times$ tadpole parameter, $u_0$) 
$m_{0l}^{asq}$ and $m_{0s}^{asq}$. 
Column 3 is the lattice spacing values in units of $r_1$ after 
`smoothing'~\cite{milcreview}. 
Column 8
gives the number of configurations and time sources per configuration 
that we used for calculating correlators. On set 5 only half the 
number were used for light quarks.}
\end{table}

On each configuration we generate and store random-wall HISQ propagators for 
several source time slices for light, strange and charm quarks:
\begin{equation}
g^{\rm HISQ}({\bf x},t_0)=M^{-1}_{x,x^\prime}\eta(t_0)_{x^\prime},
\end{equation}
where $\eta(t_0)_{x^\prime}$ is a three-component complex unit vector of 
random numbers at each site of the source timeslice, $t_0$ and zero elsewhere.

The HISQ action uses an  additional application of the fattening step
of the ASQTAD formulation, reducing discretization errors to the extent that
it is possible to simulate relativistic charm quarks on configurations of 
modest lattice spacing.

Bottom quarks are too massive to simulate relativistically on these lattices. 
However within bound states, the $b$ quark is generally slow enough 
($v_b/c\sim0.01$ in $B_c$) to treat non-relativistically. The use of the NRQCD action for b quarks is a well-developed procedure.~\cite{thackerlepage, nakhleh, oldups} 

We evolve the NRQCD propagator recursively:
\begin{equation}
G_i(x, t+1)= \left(1-\frac{\delta H}{2}\right)\left(1-\frac{H_0}{2n}\right)^nU_t
^\dagger(x)\left(1-\frac{H_0}{2n}\right)^n\left(1-\frac{\delta H}{2}\right)G_i(x
,t),
\end{equation}
with 
\begin{eqnarray}
\label{deltaH}
\delta H &=& -c_1\frac{(\Delta ^{(2)})^2}{8(M^0)^3} + c_2\frac{ig}{8(M^0)^3}
({\tilde{\Delta} \cdot \tilde{E} - \tilde{E} \cdot \tilde{\Delta}})
-c_3\frac{ig}{8(M^0)^3}{\bf\sigma}\cdot({ {\tilde{\Delta} \times \tilde{E}} - 
{\tilde{E}
 \times \tilde{\Delta}}})\nonumber\\
&&-c_4\frac{g}{2M^0}{\bf\sigma}\cdot {\bf\tilde{B}} 
+ c_5\frac{a^2\Delta^{(4)}}{24M^0}
-c_6\frac{a(\Delta^{(2)})^2}{16n(M^0)^2}.
\end{eqnarray}
The tilde expressions $\tilde{\bf E}$ and $\tilde{\bf B}$ are improved versions 
of the
naive lattice chromo-electric and chromo-magnetic fields,
${\bf E}$ and ${\bf B}$.
We use the tree-level values of $c_i=1$ for the constants.

To double statistics, we evolve the NRQCD propagator both forward and backward 
across the lattice from the source timeslice.

As we have used a random-wall source for the HISQ propagators,
it is critical that we initialize the NRQCD $b$ propagators with the 
{\em same} random-wall function $\eta(t_0)_{x^\prime}$ as we used for the 
HISQ propagators. 
This is slightly non-trivial in that the HISQ staggered fermions, and the random
wall vector $\eta(t_0)_{x^\prime}$, have one Dirac
component per site, while the NRQCD $b$ quarks have two upper and/or two lower
Dirac components. The trick is to undo the staggering transformation by 
multiplying the noise source $\eta(t_0)_{x^\prime}$ at each site with the 
four-component staggering operator:
\begin{equation}
\Omega(x)=\gamma_0^{x_0}\gamma_1^{x_1}\gamma_2^{x_2}\gamma_3^{x_3}.
\end{equation}

Furthermore, to isolate the meson ground-state, we
smear the $b$ propagator source with a Gaussian smearing function of varying 
radii $r_i$. Therefore, on timeslice $t_0$ we initialize the NRQCD
propagator as:
\begin{equation}
G^{\rm NRQCD}_i({\bf x},t_0) = \sum_{x^\prime}S(\left|x-x^\prime\right|;r_i)\eta
_{x^\prime}(t_0)\Omega(x^\prime)\Gamma,
\end{equation}
where $\Gamma$ is an element of the Dirac algebra chosen to project out a 
desired meson state.

At the sink end we must also multiply $\Omega(x)$ back into the HISQ propagator
so that we can get a multi-Dirac-component object to trace with the NRQCD
$b$ propagator:
\begin{equation}
G^{\rm HISQ}({\bf x},t)_{ab}=g^{\rm HISQ}({\bf x},t_0)_x\Omega({\bf x},t)_{ab}
\end{equation}
Then our $B$ meson correlator matrix is:
\begin{equation}
\label{B_corr}
{C_\Gamma(t-t_0)_{ij}}=\sum_{\bf x}{G^{\rm HISQ}}^\dagger({\bf x},t)\Gamma S(\left|x-x^\prime\right|;r_j)G^{\rm NRQCD}_i({\bf x^\prime},t).
\end{equation}

\section{Analysis}
We extract $B$ meson energies from the matrix of correlators (\ref{B_corr}) 
using a Bayesian factorizing fit to the form
\begin{equation}
C_{\Gamma}(t-t_0)_{ij}= \sum^{N_{\exp}}_{k=1} a_{i,k}a^*_{j,k}e^{-E_k(t-t_0)
}
+ \sum^{N_{\exp}-1}_{k^\prime=1}   b_{i,k^\prime}b^*_{j,k^\prime}(-1)^{(t-t_0)
}e^{-E^\prime_{k^\prime}(t-t_0)},
\end{equation}
where the second term fits the oscillating component inherent in staggered 
meson correlators.

We look for high-confidence fits stable with respect to varying $N_{\exp}$, and
$t_{\rm min}$ of the fit range. Where possible we simultaneously fit all the 
correlators coming from the same ensemble, to better account for correlated errors. 

In practice we fit $B$ (light) and $B_s$ together in all cases 
except the 
fine ensemble (set 5). We fit a $3\times 3$ matrix of smeared correlators in all
cases except for the $B_c$ fits on the  very-coarse ensembles (sets 1 and 2).
We always fit the pseudoscalar and vector states simultaneously.

We are interested in the ground-state energies $E_0$ and the ground-state
of the oscillating parity-partner channel $E_0^\prime$. A factor of 
$\gamma_0\gamma_5$ relates the spin structure $\Gamma$ of the direct channel
with that of the parity partner channel, $\Gamma^\prime$. In this way a measured
pseudoscalar correlator also contains a scalar meson correlator, and a 
vector correlator also contains an axial vector correlator at no extra cost. 

Because the relativistic relation between energy and mass does not hold for 
NRQCD $b$ quarks, there is an unknown energy shift between the physical masses 
we are interested in and the the fitted energies.  Instead we measure the 
splitting between the state of interest and a similar state 
with the same NRQCD quark content. 

We convert this splitting to physical units using 
$r_1=0.3133(23)$fm \cite{Davies:2009ts},  giving a 0.7\% uncertainty in any 
measured splitting in our lattice calculation. Hence we can minimize the 
scale-setting error by 
choosing comparison states as close as possible to the state of interest.
We are perfectly free to construct a fictitious comparison state which is a
composite of real states, provided all components have well-known experimental
and lattice measurement for calibration.

We consider three methods to determine the $B_s$ and $B_c$ masses: 
\romeqn
\setcounter{equation}{0}
\begin{equation}
\label{EXPI}
M_{B_{s/c}} = \left(E_{B_{s/c}} -\frac{1}{2}E_{b\overline{b}}\right)_{\rm latt} 
+ \frac{1}{2}M_{b\overline{b}}
\end{equation}
\begin{equation}
\label{EXPII}
M_{B_c} = \left(E_{B_c} -\frac{1}{2}(E_{b\overline{b}}
+E_{c\overline{c}})\right)_{\rm latt} 
+ \frac{1}{2}\left(M_{b\overline{b}} + M_{c\overline{c}}\right)
\end{equation}
\begin{equation}
\label{EXPIII}
M_{B_c} = \left(E_{B_c} - (E_{B_s}+E_{D_s} - E_{\eta_s})\right)_{\rm latt} 
+ \left(M_{B_s} + M_{D_s} - M_{\eta_s}\right)
\end{equation}
\reseteqn
\renewcommand\theequation{\thesection.\arabic{equation}}

Here $E_{b\overline{b}}$, for example, refers to the spin-averaged lattice energy
of $b\overline{b}$ states. In each equation we must apply the lattice scale
$a^{-1}$ (and its uncertainty) to the expression in the $\left(\right)_{\rm latt}$ only.

Where the subtraction  compares 
states with different electromagnetic charge structures we must estimate  
the adjustment necessary to account for 
electromagnetic effects.
\section{Results and discussion}
\subsection{Pseudoscalar states}
In practice Method \ref{EXPI} is the only one applicable to $B_s$ spectroscopy.
We extract the lattice energies of the $B_s$ states from each of the ensembles,
convert to physical masses via expression \ref{EXPI}. 

\begin{figure}
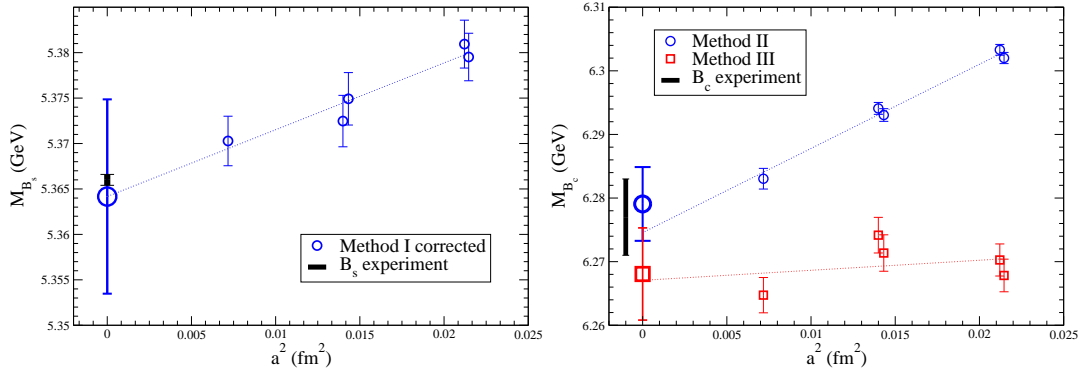

\begin{center}
\scalebox{0.28}{\includegraphics{GRAPHICS/MBs.methods.I.extrap.10nov09.shifted.allerr.eps}}
\scalebox{0.28}{\includegraphics{GRAPHICS/MBc.all_methods_extrap.09nov09.eps}}
\end{center}
\caption{\label{MBs_plot}
Lattice calculations for $M_{B_s}$ (left) $M_{B_c}$ (right) and  with 
energy shift subtracted and resulting masses extrapolated in $a^2$ to the 
physical point. 
For $M_{B_s}$ we correct the finite-$a^2$ points for $b$ and $s$ mass mistuning
before extrapolation.
For $M_{B_c}$ the 
continuum points are corrected upwards for electromagnetic effects by
4.5MeV and 1MeV for Methods I and II respectively.
Error bars on extrapolated points reflect total errors.}
\end{figure}

After the recent, more precise determination of $r_1$ \cite{Davies:2009ts},
it has become apparent that both the $s$ quark mass and the $b$ quark mass
were tuned too high. Method I for $M_{B_s}$ is particularly sensitive
to the mistuned quarks. We have estimated the effect of the 
mistuned quark masses by substituting into Method \ref{EXPI} mesons with 
different
valence masses. We estimate that for the very coarse, coarse and fine 
ensembles, the too-large strange mass pushes up $M_{B_s}$ by 7.5, 10 and 9MeV, 
respectively. The $b$ mistunings bias $M_{B_s}$ up by 10.5, 13, and 15MeV
on the same ensembles. We correct for these biases in the finite $a$ 
calculations, and then extrapolate, estimating an additional systematic 
uncertainty of 10MeV on the extrapolated value, giving:
\begin{equation}
M_{B_s}=5.341(4)(10){\rm GeV},
\end{equation}
with the first error being statistical and the second, dominant, error being
the quark-tuning systematic error. Figure \ref{MBs_plot} (left) illustrates
the extrapolation.

For $B_c$ pseudoscalars we can use Methods \ref{EXPII} and \ref{EXPIII}. As 
\ref{EXPII} is superior to \ref{EXPI} and they are not linearly independent
we do not also consider \ref{EXPI} here. We again extrapolate in $a^2$ to the
continuum for each.

We correct for the electromagnetic structure mismatch.
Method \ref{EXPII} compares neutral 
$b\overline{b}$ and $c\overline{c}$ states with the charged $B_c$ state. We 
calculate that this mismatch causes an underestimate of $M_{B_c}$ 
by $\sim 4.5\pm 2$MeV.
In Method \ref{EXPIII}, comparing similarly charged $B_c$ and $D_s$ introduces 
an underestimate of $\sim 1\pm 1$MeV. 

After correcting the electromagnetic contribution we get:
\begin{equation}
M_{B_c}(II) = 6.279(2)(1)(5)(2){\rm GeV}
\end{equation}
\begin{equation}
M_{B_c}(III) = 6.268(4)(6)(1)(1){\rm GeV},
\end{equation}
where the errors are (statistical)($r_1$)(NRQCD)(EM). The agreement between 
the two independent subtraction methods is a strong test of our control
of systematics. Because the $()_{\rm latt}$ term is very small, both are 
quite insensitive to $b$ and $s$ tuning, and no further subtraction is 
necessary. Results from both methods are in excellent 
agreement with the PDG average of $6.277(6)$ GeV\cite{pdg09}.
See Figure \ref{MBs_plot}, right. 

The HISQ $c$ quark seems to be the source of the 
strong discretization effects in Method II, which go as 
$\alpha_s(v/c)^2(am_c)^2$. The $c$ quark is more relativistic inside the
$B_c$ than in a $c\overline{c}$, so these errors do not cancel exactly, but
should vanish in the continuum.

\subsection{Vector states}
For the $B^*$ states there is an obvious method of correcting for the NRQCD
energy shift --- compare to the nearby $B$ pseudoscalar states to get the 
hyperfine splitting. The remaining complication is that since the 
$\sigma\cdot B$ term in the NRQCD action generates the $B^*-B$
splittings, radiative corrections to this term could generate a multiplicative
correction to the splitting. (Recall we have used the tree-level $c_4$.)
We therefore use the hyperfine splitting of the $B_s$ system as calibration
for that of the $B_c$ system and calculate:
\begin{equation}
R_c=\frac{E_{B_c^*}-E_{B_c}}{E_{B_s^*}-E_{B_s}},
\label{ratio}
\end{equation}
which will cancel all of the NRQCD energy shifts, multiplicative corrections, 
and scale-setting error. We extrapolate to $a^2=0$, multiply by the PDG
average  value of $M_{B_s^*}-M_{B_s}=46.1(1.5)$MeV~\cite{pdg09}, and
add the experimental $B_c$ mass, giving us a {\em prediction}
of the $B_c^*$ mass of $M_{B_c^*}=6.330(7)(2)(6)$ GeV.
As a check we also calculate $R_l$ with light quark $B$ and $B^*$ states.
A complete discussion of the $B_c$ hyperfine splitting calculation can be found in \cite{Gregory:2009hq}.

\begin{figure}
\begin{center}
\scalebox{0.38}{\includegraphics{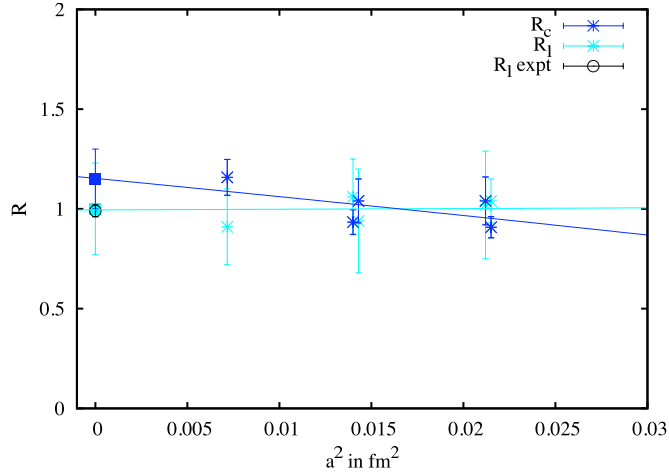}}
\end{center}
\caption{\label{split_ratio}
 The extrapolation of the ratio $R$ to the continuum for both $B_c^*$ and $B^*$.}
\end{figure}

\subsection{Scalar states}
As mentioned in Section 3, the oscillating component of the 
pseudoscalar correlators gives us the scalar  states. 

\begin{figure}
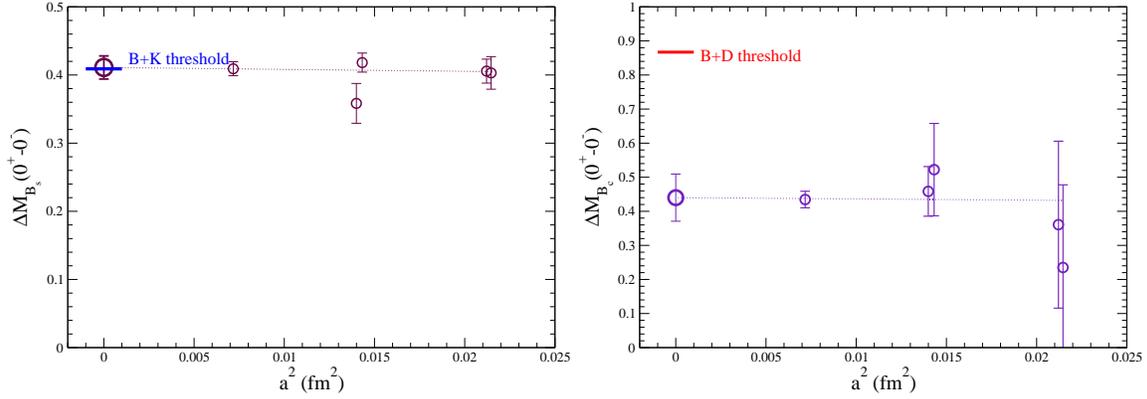

\begin{center}
\scalebox{0.3}{\includegraphics{GRAPHICS/Bs.EoEsplit.sca-ps.09nov09.eps}}
\scalebox{0.3}{\includegraphics{GRAPHICS/Bc.EoEsplit.sca-ps.09nov09.eps}}
\end{center}
\caption{\label{scalars} The splitting between the scalar and pseudoscalar 
states for $B_s$ (left) and $B_c$ (right). Values are extrapolated to the 
continuum. Shown on each are the relevant hadronic thresholds.}
\end{figure}

We extract the $E_{0^+}-E_{\rm 0^-}$ splittings directly in the simultaneous 
fits. 
Converting to physical units we again extrapolate in $a^2$ to the 
continuum and find:
\begin{equation}
\Delta M_{B_s}(0^+-0^-) = 0.41(2){\rm GeV}
\end{equation}
\begin{equation}
\Delta M_{B_c}(0^+-0^-) = 0.44(7){\rm GeV},
\end{equation}
quoting statistical errors only. 
As this splitting is generated by the kinetic term, it should acquire no
multiplicative renormalization, and most systematics should cancel.
The $O^+$ $B_c$ state lies about $400$MeV below the $B+D$ 
threshold so it should be a narrow state. It is less clear whether the $0^+$
$B_s$ state is below the $B+K$ state, but in any case it should be close 
enough to the $B+K$ that it should also be a narrow state. See Figure 
\ref{scalars}, right and left, respectively.

\section{Conclusions}
We have shown preliminary results of precise lattice calculations of 
pseudoscalar masses in the $B_s$ and $B_c$ system, and of 
vector-pseudoscalar and scalar-pseudoscalar splittings. 
Our calculation of $M_{B_s}$ and $M_{B_c}$ agree within errors with 
experimental measurements of these states. Some work remains to fully 
understand the systematic errors and biases related to  mistuning of 
quark masses.

Our calculations of the $B_s$ and $B_c$ scalars and the $B_c^*$ vector constitute
{\em predictions} of the masses of these states before experimental measurement.

The precision and accuracy of these results reaffirms that the combination
of HISQ light quarks and NRQCD $b$ quarks is a powerful lattice technique.
Further work will complete the exploration of the lowest  $B$,$B_s$ and $B_c$
states, and then apply these techniques to the calculation of form-factors 
and decay constants relevant to weak-matrix elements.

\end{document}